\documentclass[pre,twocolumn,showpacs,amsmath,amssymb]{revtex4}
\usepackage{graphicx}% Include figure files
\usepackage{bm}% bold math
\usepackage{amssymb}
\begin{document}

\title{Particle simulation of vibrated gas-fluidized beds
of cohesive fine powders}
\author{Sung Joon Moon, I. G. Kevrekidis, and
S. Sundaresan\footnote{Corresponding author:
sundar@princeton.edu; 609-258-4583 (tel); 609-258-0211 (fax).}}
\affiliation{Department of Chemical Engineering \&
Program in Applied and Computational Mathematics\\
Princeton University, Princeton, NJ 08544}

\begin{abstract}
We use three-dimensional particle dynamics simulations,
coupled with volume-averaged gas phase hydrodynamics,
to study vertically vibrated gas-fluidized beds of fine,
cohesive powders.
The volume-averaged interstitial gas flow is restricted
to be one-dimensional (1D).
This simplified model captures the spontaneous development
of 1D traveling waves, which corresponds to bubble formation
in real fluidized beds.
We use this model to probe the manner in which vibration and
gas flow combine to influence the dynamics of cohesive
particles.
We find that as the gas flow rate increases, cyclic pressure
pulsation produced by vibration becomes more and more significant
than direct impact, and in a fully fluidized bed this pulsation
is virtually the only relevant mechanism.
We demonstrate that vibration assists fluidization by creating
large tensile stresses during transient periods, which helps
break up the cohesive assembly into agglomerates.
\end{abstract}

%\vspace*{0.4in}
%{\bf key words:} vibrated bed, cohesive particles, particle simulation,
%pressure pulsation, fluidization, tensile strength.\\

\maketitle

\section{Introduction}

It is well known that fine, cohesive particles cannot
be fluidized easily.~\cite{geldart73,visser89}
When fluidization by an upflow of gas is attempted,
assemblies of such particles tend to lift up as a plug
and only form ratholes and cracks through which the fluid
escapes. Attractive inter-particle forces frequently arise as
a result of capillary liquid bridges or van der Waals forces.
A variety of techniques to achieve smooth fluidization of such
particles have been explored in the literature. When the cohesion
arises from van der Waals forces, as in the case of Geldart
type C particles which are typically 30 $\mu$m or smaller in
size,~\cite{geldart73} coating with hard nanoparticles~\cite{mohan04}
or an ultrathin film~\cite{ferguson00} can weaken the attraction
between the bed particles, thus enabling smooth fluidization.
Alternate approaches to facilitate fluidization include causing
agglomerate break-up through a secondary supply of energy using
mechanical vibration,~\cite{dutta91,jaraiz92,marring94,valverde01,
wank01,mawatari02, mawatari03,nam04} acoustic
waves~\cite{chirone93,nowak93,russo95} or an oscillating magnetic
field.~\cite{yu05}
Such approaches have been shown to be effective even for beds
of nanoparticles.~\cite{nam04}
In the present study, we are concerned with some aspects of the
mechanics of vibrated fluidized beds of cohesive particles.
	
	Vibrated layers of granular materials have been
studied extensively in the literature and formation of
patterns in shallow granular layers is now well
known.~\cite{melo94,umbanhowar96}  Mixing and segregation in
such vibrated beds have also received considerable
attention.~\cite{moebius01,burtally02,moon03} The motion of
large intruders in vibrated beds, leading to the well-known
Brazil nut and reverse Brazil nut effects,~\cite{moebius01,hong01}
has also received much attention in the literature,
where experimental measurements suggest a non-negligible influence
of the interstitial gas on the observed flow
patterns.~\cite{naylor03,moebius05}
In general, the deeper the bed and/or the smaller the bed particles,
the greater the influence of the interstitial gas phase on the
dynamics of the assembly of particles. 

	In vibrated fluidized beds, where vibration is supplemented
with a fluidizing gas flow (or {\it vice versa}), the importance of
the interstitial gas is obvious as the drag due to the gas
flow supports a substantial portion of the weight of the particles.
Vibrated fluidized beds have found many applications in industrial
practice (e.g., see an article by Squires~\cite{squires04}).
Understanding the manner
in which the vibration aids the fluidization process is important
both for macroscopic analysis of vibrated fluidized beds and for
detailed interrogation of agglomerate size distribution and mixing
at the agglomerate and particle scales. 

	Predicting the minimum fluidization velocity of a vibrated
fluidized bed (of cohesive particles) is perhaps the simplest
quantitative, macroscopic
analysis problem one can think of. This indeed has been a subject
of many investigations~\cite{musters77,liss83,erdesz86,marring94,
tasirin01,mawatari02,nam04,wank01} and different approaches have
been proposed in the literature to incorporate the effect of
vibration on the overall force balance used to determine the
minimum fluidization velocity. 
Musters and Rietema~\cite{musters77} suggested that additional
terms need to be included in the force balance relation
in order to account for the increased pressure drop due to
cohesion. They included additional terms for inter-particle
forces as well as wall friction.
Liss {\it et al.}~\cite{liss83} proposed an additional term
to account for the effect of cohesion arising from liquid bridges.
Wank {\it et al.}~\cite{wank01} showed that the agglomerate
size decreases as the vibration intensity increases, and
studied the effect of the pressure on the minimum fluidization
velocity.
Erd\'esz and Mujumdar~\cite{erdesz86} developed a theory
which includes the effect of vibration in the prediction of
the minimum fluidization velocity;
they found that the pressure drop decreased with increasing
vibration intensity in their experiments with various particles
in the range 0.15-2.75 mm.
However, Tarisin and Anuar~\cite{tasirin01} found the opposite
trend in their study of vibrofluidization of particles of
15 $\mu$m to 34 $\mu$m, and others~\cite{marring94,mawatari02,nam04}
found no appreciable dependence on the vibration intensity with
particles of range 12 nm through 100 $\mu$m.

	Such conflicting reports stem from our limited understanding
of the manner in which the vibrating boundary interacts with the
assembly of particles in the bed. One can readily envision at least
two modes of interaction: (a) When the effect of the interstitial
gas on particle motion is of negligible importance, the vibrating
base plate clearly imparts impulse to the particle assembly
periodically only through direct collisions. (b) When gas-particle
drag is non-negligible, a periodically varying pressure field can
be expected to develop in the bed, and the base plate interacts
with the bed particles indirectly by driving these pressure
pulsations as well. The relative importance of these two
contributions can be expected to depend on the bed depth,
particle size, and the superficial velocity of the fluidizing
gas; however, quantitative estimates are unavailable.

At a more detailed level, it is readily apparent that vibration
results in the formation of small agglomerates which are more
amenable to fluidization, yet how such small agglomerates are
formed is not understood. A clear understanding of this
mechanism is an important first step in any effort to capture
the effect of vibration on the dynamics of agglomerates
(e.g., in a population-balance-type model for vibrated
fluidized beds). At first glance, one may speculate that
vibration drives vigorous collisions between agglomerates,
which in turn causes their break-up; but it is not known if
this is indeed the dominant mechanism.

	In the present study, we have examined the interaction
of the vibrating base plate with a bed of particles through
detailed simulations. We will demonstrate that as the fluidizing
gas velocity is increased, the interaction with the boundary
through pressure pulsation becomes more and more dominant and
that above the minimum fluidization conditions, the interaction
occurs almost exclusively through pressure pulsations.
We will also show that the large tensile stress induced by the
vibration is a more important mechanism (than vigorous
collisions between agglomerates) in causing the break-up of
the cohesive particle assembly.

	The dynamic behavior of gas fluidized beds containing
a large number of particles has been widely examined in the
literature through two-fluid models.~\cite{gidaspow94,fan98,jackson00}
This approach requires
closure relations for gas-particle interactions and the stresses.
Considerable progress has been made in developing and validating
the closures for assemblies of non-cohesive
particles,~\cite{gidaspow94,fan98,jackson00} but not for the
case of cohesive materials. Accurate two-fluid model
boundary conditions that capture the effect of vibrating
boundaries are also unavailable. Furthermore, the two-fluid
model approach is not well suited for investigation of the
mechanism of agglomerate break-up.
For these reasons, we model vertically vibrated gas fluidized
beds of fine powders using a hybrid scheme, where the solid phase
is treated as discrete spherical particles, following the so-called
discrete element method (DEM; or soft-sphere molecular
dynamics),~\cite{cundall79} while the gas phase is treated as
continuum, following the volume-averaged hydrodynamics
just like in a two-fluid model.~\cite{gidaspow94,jackson00}
This DEM-based hybrid approach was proposed by Tsuji
{\it et al.}~\cite{tsuji93} in their simulation of two-dimensional
(2D) fluidized beds, and it has been subsequently refined by 
others.~\cite{hoomans96,xu97}
Recently, this model has been used to study fluidized beds of
Geldart A particles,~\cite{ye04} the effect of an arbitrary
cohesive force proportional to the particle buoyant
weight,~\cite{rhodes01a,rhodes01b} and segregation in fluidized
beds of bidisperse particles.~\cite{bokkers04}

	Even with the high speed computing available today,
one can only simulate small systems, which are orders of
magnitude smaller than real vibrated fluidized beds, through
this hybrid approach. Therefore, it is essential that one
choose suitable, idealized problems to probe the underlying
mechanics. With this in mind, we have chosen essentially
one-dimensional (1D) vibrated fluidized beds of particles,
applying periodic boundary conditions in the two lateral
directions. This limits the macroscopic dynamics to the
vertical direction only, and one obtains one-dimensional
traveling waves (1D-TW) instead of bubble-like voids observed
in experiments and in fully three-dimensional flow simulations;
nevertheless, such an idealized problem is, in our opinion,
adequate to expose the manner in which the vibrating base
plate interacts with the bed particles and how dense cohesive
regions are broken down into smaller agglomerates. 

The rest of the paper is organized as follows.
The DEM-based hybrid model is described in Sec.~\ref{method},
and bubble formation, realized as 1D-TW in our geometry
(narrow cross sectional areas), is presented in Sec.~\ref{1DTW}.
The effect of cohesion and vibration on the fluidization is
presented in Sec.~\ref{cohesion},
and the manner in which the vibration enhances the fluidization
will be discussed in Sec.~\ref{vibration}.
The pressure drop in vibrated fluidized beds and the mechanism
through which the vibration breaks up cohesive assemblies will
be presented in Sec.~\ref{pressuredrop} and Sec.~\ref{breakup}
respectively,
which are followed by the conclusions in Sec.~\ref{conclusion}. 

\section{Method: DEM-based hybrid model}
 \label{method}

Since its introduction by Cundall and Strack~\cite{cundall79}
nearly three decades ago, the DEM has been successfully used
in modeling various particulate flow problems,
including hopper flows,~\cite{zhu04} shearing
cells,~\cite{volfson03} rotating drums,~\cite{mccarthy96}
and oscillated layers.~\cite{mehta91,wassgren97}
Comprehensive description of this method can be found in the
literature.~\cite{herrmann98,duran00,rapaport04}
Below, we briefly describe the main idea of the DEM, and
subsequently focus on how the volume-averaged gas phase
hydrodynamics is coupled with the individual particle
dynamics in our model.

\subsection{Discrete element method}
 \label{DEM}

In the DEM simulation, particles are modeled as ``soft''
sphere (the deformation is accounted for by overlaps),
whose trajectories are computed by integrating
Newton's equations of motion.
When objects (particles or system boundaries) get into contact,
the interaction is resolved by decomposing the interaction force
into the normal and tangential directions relative to the
displacement vector between the objects at contact,
$\mathbf{F}_{cont} = ({\mathbf F}_n,{\mathbf F}_s)$,
and the energy dissipation upon contact is characterized by
the inelasticity and the surface friction.
We use the so-called spring-dashpot model, with a Hookean
spring.~\cite{cundall79} The objects are allowed to overlap
upon contact, and the contact force in the normal direction
${\mathbf F}_n$ is determined by the amount of overlap $\Delta_n$
and the normal component of the relative velocity at contact $v_n$,
\begin{equation}
 \label{normal}
{\mathbf F}_n = (k_n \Delta_n - \gamma_n v_n) \hat{\mathbf n},
\end{equation}
where $k_n$ is the spring stiffness in the normal direction,
$\gamma_n$ is the damping coefficient, and $\hat{\mathbf n}$
is the unit vector in the normal direction at contact,
pointing from the contact point toward the particle center.
The damping coefficient $\gamma_n$ is related to $k_n$ by the normal
coefficient of restitution $e$ ($0 \leq e \leq 1$),
\begin{equation}
 \label{restitution_coeff}
\frac{4k_n/m^*}{(\gamma_n/m^*)^2} = 1 + \left(\frac{\pi}{\log e}\right)^2,
%\frac{\gamma_n}{m^*} = \sqrt{\frac{4k_n/m^{*2}}{1+(\pi/\log e)^2}},
\end{equation}
where $1/m^* = 1/m_i + 1/m_j$, and $i$ and $j$ are indices
of interacting particles or objects.
In principle, the value of $k_n$ is determined by Young's modulus
of the material under consideration.
However, unless stated otherwise, we use a much smaller value for
$k_n$, compared to the one computed based on the usual range of
Young's modulus.
If the main results are not qualitatively different, it is favorable
to use a smaller value of $k_n$, because the collision duration
time in DEM scales with $k_n^{-1/2}$, which determines the
integration time-step size required to accurately resolve the
interaction during the contact.
This is a well-known issue in DEM simulations.~\cite{herrmann98,duran00}
We varied $k_n$ over three orders of magnitude and verified that
the main results do not depend sensitively on the choice of $k_n$,
even though actual contact force between the objects certainly
depend on the value of $k_n$.
For instance, the results for two $k_n$ values differing by a
factor of 10 will be presented in Figs.~\ref{platestress},
\ref{platestress2}, and \ref{pdropAVG}.

The interaction in the tangential direction is modeled by a
``spring and slider'', and the contact force is given by:
\begin{equation}
 \label{tangential}
{\mathbf F}_s = -{\rm sign}(v_s)\times{\rm min(}k_t \Delta_s, \mu |{\mathbf F}_n|) \hat{\mathbf s},
\end{equation}
where $v_s = {\mathbf v}_s\cdot\hat{\mathbf s}$ is the tangential
component of the relative velocity at contact; ${\mathbf v}_s
= \hat{\mathbf n}\times({\mathbf v}_{ij}\times\hat{\mathbf n})$;
$\hat{\mathbf s}$ is the unit vector in the tangent plane
collinear with the component of the relative velocity at
contact; $k_t$ is the tangential spring stiffness that is
related to $k_n$ by the Poisson's ratio of the material $\nu_P$
[$k_t = 2k_n(1-\nu_P)/(2-\nu_P)$];
and $\Delta_s$ is the magnitude of tangential
displacement from the initial contact.
The magnitude of the total tangential force is limited by the
Coulomb frictional force $\mu |{\mathbf F}_n|$, where $\mu$ is
the coefficient of friction.
More sophisticated and/or realistic interaction models, such as that
of Walton and Braun's~\cite{walton86} or a Hertzian spring-dashpot
model,~\cite{schafer96} may also be used.
However, we choose a simple Hookean spring-dashpot and spring-slider
model, as it has been shown to successfully reproduce many
experimental observations,~\cite{zhu04,mccarthy96,wassgren97}
and it is computationally more tractable than others.

Among different inter-particle forces, other than due to contact,
we consider only cohesion arising from van der Waals force.
In principle, the cohesion can depend on the particle characteristics,
such as polarizability, particle size, and asperity.~\cite{israel97}
However, we adopt a simple formula by Hamaker, as we aim to bring
out the effect of cohesion on the fluidization behavior,
rather than to validate different cohesion models.
Particulate flows in industry often consist of particles with a wide
range of sizes and shapes; however, we seek a better understanding of
simple systems consisting of monodisperse spheres, which are
well characterized by a small set of parameters.
The cohesive van der Waals force between two spheres of radii
$r_i$ and $r_j$  can be expressed as,~\cite{israel97}
\begin{eqnarray}
 \label{hamaker}
\mathbf{F}_c &=&  - \frac{A}{3} \frac{2r_ir_j(s+r_i+r_j)}{[s(s+2r_i+2r_j)]^2}
\times \nonumber \\
&& \left[\frac{s(s+2r_i+2r_j)}{(s+r_i+r_j)^2-(r_i-r_j)^2} -1 \right]^2
\hat{\mathbf{n}} \nonumber \\
&\approx& - \frac{A}{12}\frac{r}{s^2}\hat{\mathbf{n}}
~~~
{\rm (for~} r = r_i = r_j~~{\rm and~} s \ll r{\rm )},
\end{eqnarray}
where $A$ is the Hamaker constant, and $s$ is the inter-surface distance.
As the original formula is a rapidly decreasing function of $s$,
further simplification using the assumption of $s \ll r$ has been made.
This model has a singularity at contact.
In order to avoid this artifact, we introduce a widely accepted
minimum cut-off value for the inter-surface distance of 0.4 nm
($\equiv \delta^*$), which corresponds to the inter-molecular
center-to-center distance.~\cite{seville00}
In what follows, the level of cohesion is represented by the cohesive
Bond number $Bo$, which is defined as the ratio of the maximum cohesive
force (at the minimum cut-off separation $\delta^*$) to the particle weight.
Other types of cohesion can be readily accounted for in DEM-based
models.~\cite{mikami98,rhodes01b,kuwagi02}
%the model.
%For instance, the cohesion arising from capillary forces has
%been incorporated in DEM-based simulation to study its effects on
%fluidization

\subsection{Coupling with gas phase hydrodynamics}
 \label{coupling}

The dynamics of individual particles is coupled with the
volume-averaged gas phase hydrodynamics.~\cite{tsuji93}
In this hybrid model, the equations of motion for individual particles
have two additional terms (compared to traditional DEM modeling particles
in vacuum) arising from the presence of the gas phase:
\begin{equation}
\label{eachgrain}
m_p\frac{d\mathbf{v}_p}{dt} = m_p\mathbf{g}_{eff} + \mathbf{F}_{cont} +
\mathbf{F}_c + \frac{V_p}{\phi}\beta(\phi)\left(\mathbf{u}_g -
\mathbf{v}_p \right)- V_p \nabla p,
\end{equation}
where $m_p$ and $\mathbf{v}_p$ are individual particle mass and
velocity, respectively.
The first term on the right hand side represents the body force due
to gravity; $\mathbf{g}_{eff}$ is the effective gravitational
acceleration in the reference frame where equations are integrated.
For non-vibrated beds, $\mathbf{g}_{eff}$ is simply the gravitational
acceleration $\mathbf{g}$.
When the bed is subject to a single frequency oscillation,
the equations are integrated in the vibrated frame, and
$\mathbf{g}_{eff} = \mathbf{g}[1 + \Gamma\sin(2\pi ft)]$,
where $\Gamma = A_p(2\pi f)^2/g$ is the maximum acceleration
of the base plate (distributor) non-dimensionalized by the gravitational
acceleration $g = |\mathbf{g}|$, $A_p$ is the oscillation amplitude,
and $f$ is the oscillation frequency.
We assume the oscillating base plate is made of the same materials
as the particles (the same values for $e$ and $\mu$),
and that the mass of the plate is infinitely large compared to that
of an individual particle.
The second term and the third term represent the aforementioned
contact force and van der Waals force, respectively.
The total force acting on the particles due to the fluid is commonly
partitioned into the local drag part and the effective buoyant part,
as was done here (see e.g., an article by Ye et al.~\cite{ye04}):
The fourth term accounts for the drag force, and the last term
accounts for the contribution of the gradually varying part of the
pressure field,
where $V_p$ is the volume of each particle; $\phi$ and
$\mathbf{u}_g$ are volume-averaged solid phase volume fraction and
gas phase velocity, respectively; $\beta$ is the inter-phase
momentum transfer coefficient~\cite{gidaspow94};
and $p$ is the gas phase pressure.

In general, the gas phase quantities are obtained by simultaneously
integrating the coarse-grained mass and momentum balance equations.
We assume the gas phase to be incompressible, which will be validated
later (see Sec.~\ref{pressuredrop}).
The addition of continuity equations for the gas phase and solid
phase reads
\begin{equation}
\label{gas_cont}
\nabla \cdot \left[ (1 - \phi) \mathbf{u}_g + \phi \mathbf{u}_s \right] = 0,
\end{equation}
and a reduced momentum balance equation for the gas phase, based
on generalized Darcy's law, is given as
\begin{equation}
\label{simpleNS}
0 =  -(1 - \phi)\nabla p + \beta(\phi) \left(\mathbf{u}_s - \mathbf{u}_g\right),
\end{equation}
where $\mathbf{u}_s$ is the coarse-grained solid phase velocity.
Coarse-grained variables are considered only on grids
where the continuum balance equations are solved.
Note that the solid phase continuum (or coarse-grained) variables
are explicitly available in the course of DEM computation.  

\subsection{Beds of narrow cross sectional area}
 \label{1d-bed}

Only beds of narrow cross sectional areas will be considered,
and the volume-averaged gas phase (hence the solid phase coarse-grained
variables as well) is assumed to be 1D. However, the solid phase
is maintained to be 3D, as the way particles pack and collide in lower
dimensions are considerably different from those in realistic 3D cases.
Our assumption allows us to consider relatively deep beds (through
inexpensive computational effort) and to bring out the basic
physics of more complicated dynamics in higher dimensions.

Solid phase coarse-grained variables at 1D discrete grid points
are computed by distributing the particle mass and momenta to
the nearest two grid points using a halo function $h$ that
continuously decreases to zero around the particle;
\begin{equation}
h(z-z_0) = \left\{ \begin{array}{ll}
                1-{|z-z_0|/ \Delta z} &~~~ {\rm for~} |z-z_0| < \Delta z,\\
                0 &~~~ {\rm otherwise,}
                 \end{array}
        \right.
\end{equation}
where $z$ is the particle position in the vertical direction, $z_0$
is that of a neighboring grid point, and $\Delta z$ is the grid spacing.
It is readily seen that $h$ has the property that the particle
quantities are distributed to the two nearby grid points,
inversely proportional to the distance to the grid point.
The coarse-grained variables, the number density $n$ and
$\mathbf{u}_s$, on the grids are then defined simply as
\begin{eqnarray}
n(z_0) & = & \sum_{i= 1}^N h(z_i-z_0),\\
n(z_0){\mathbf u}_s(z_0) & = & \sum_{i= 1}^N h(z_i-z_0){\mathbf v}_{p,i}.
\end{eqnarray}
where $z_i$ and $z_0$ are the $i$th particle location and nearby
grid location, respectively.

The assumption of the gas phase to be 1D facilitates further
mathematical simplifications of the above particle-gas interaction
formulation.
In 1D continuum cases, Eq.~(\ref{gas_cont}) can be integrated
\begin{equation}
\label{1Dcont}
(1 - \phi) \mathbf{u}_g + \phi \mathbf{u}_s = \mathbf{U}_s,
\end{equation}
where $\mathbf{U}_s$ is the superficial gas flow velocity.
Strictly speaking, in a vibrated fluidized bed, ${\mathbf U}_s$
may also vary periodically.
The extent of its variation will depend on the dynamics of the
gas in the plenum and the flow resistance offered by distributor
(base plate). One can show that the temporal variation of
${\mathbf U}_s$ will be small for a highly resistive distributor
plate, which we assume.
Thus, in modeling of both non-vibrated and vibrated fluidized beds,
${\mathbf U}_s$ will be considered to be a time-independent
parameter.

After some manipulation, Eq.~(\ref{eachgrain}) can be rewritten
as follows:
\begin{eqnarray}
\label{oneDeq}
m_p\frac{d\mathbf{v}_p}{dt} = && m_p\mathbf{g}_{eff} + \mathbf{F}_{cont} + \mathbf{F}_c + \frac{V_p}{\phi}\beta(\phi) \times \nonumber \\
&& \left[(\mathbf{u}_s - \mathbf{v}_p) - \frac{1}{(1 - \phi)^2}(\mathbf{u}_s - \mathbf{U}_s) \right].
\end{eqnarray}
Note that the presence of the gas phase is realized as additional
terms involving coarse-grained variables, instead of separate
continuum equations to be integrated simultaneously.
In the course of integration, $\phi$ and $\mathbf{u}_s$ in
Eq.~(\ref{oneDeq}) need to be evaluated at the particle location,
not at the grid points.
We evaluate them by linearly interpolating those values at
the neighboring grid points.

For the inter-phase momentum transfer coefficient $\beta$, we use
an expression proposed by Wen and Yu~\cite{wen66}:
%who extended the work of Richardson and Zaki (1954):
\begin{equation}
\beta = \frac{3}{4} C_D \frac{\rho_g \phi(1 - \phi) |\mathbf{u}_g - \mathbf{u}_s|}{d_p} (1 - \phi)^{-2.65},
\end{equation}
where $C_D$ is the drag coefficient, $\rho_g$ is the gas phase
mass density, and $d_p$ is the particle diameter.
The drag coefficient proposed by Rowe~\cite{rowe61} is employed
in our model:
\begin{equation}
C_D = \left\{ \begin{array}{ll}
\frac{24}{Re_g}\left(1 + 0.15Re_g^{0.687}\right), & Re_g < 1000, \\
0.44, & Re_g \geq 1000,
\end{array}
      \right.
\end{equation}
where 
\begin{equation}
Re_g = \frac{(1 - \phi)\rho_g d_p |\mathbf{u}_g - \mathbf{u}_s|}{\mu_g},
\end{equation}
and $\mu_g$ is the gas phase viscosity.
As we consider fine powders, $Re_g$ is generally small, and we use
the assumption $Re_g \ll 1$, which further simplifies $\beta$:
\begin{equation}
\label{beta}
\beta(\phi) = 18 \frac{\mu_g}{d_p^2} \phi (1 - \phi)^{-2.65}.
\end{equation}

Casting Eqs.~(\ref{oneDeq}) and (\ref{beta}) in a dimensionless form,
using $\rho_s$, $d_p$, $\sqrt{gd_p}$, $\sqrt{d_p/g}$ as characteristic
density, length, velocity, and time, one obtains the following
non-dimensional groups (arrows indicate changes in the notation
from dimensional variables to non-dimensional variables that will be
used henceforth):
\begin{eqnarray*}
k_n &\leftarrow& \frac{k_n}{\rho_s g d_p^2}, ~~~~~~~~~~{\rm spring~stiffness}\cr
U_s   &\leftarrow& \frac{U_s}{\sqrt{g d_p}}, ~~~~~~~~~~{\rm superficial~gas~flow~rate}\cr
\delta   &\equiv& \frac{\delta^*}{d_p}, ~~~~~~~~~~~~~~{\rm scaled~minimum~separation~distance}\cr
Bo &\equiv& \frac{A}{4\pi \rho_s g d_p^2\delta^{*2}}, ~~~{\rm cohesive~Bond~number}\cr
St &\equiv& \frac{\rho_s g^{1/2} d_p^{3/2}}{\mu_g}, ~~~~{\rm Stokes~number}
\end{eqnarray*}
together with non-dimensional parameters, namely $\Gamma$,
$f \leftarrow f\sqrt{d_p/g}$, $e$, $\mu$, and $\nu_P$.

\begin{figure}[t]
\begin{center}
\includegraphics[width=.8\columnwidth]{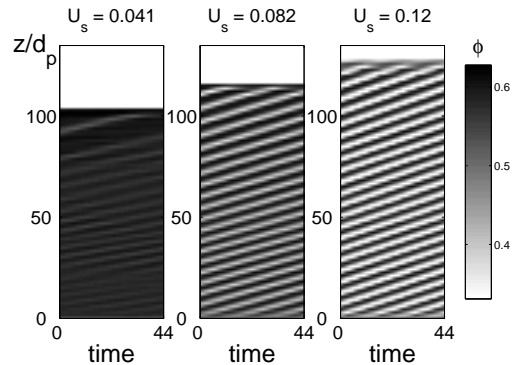}
\end{center}
\caption{\label{Usthree}
Spatiotemporal plots of conventional fluidized beds of non-cohesive
particles ($Bo = 0.0, St = 55$) at three different
superficial gas flow rates.
Gray scale represents the local volume fraction of particles;
both the regions of the minimum value ($\phi \sim 0.33$) in the
gray scale bar and completely void regions ($\phi = 0$) above
the bed top surface are shown in the same white color.
}
\end{figure}

\section{Results and Discussion}

\begin{figure}[t]
\begin{center}
\includegraphics[width=.8\columnwidth]{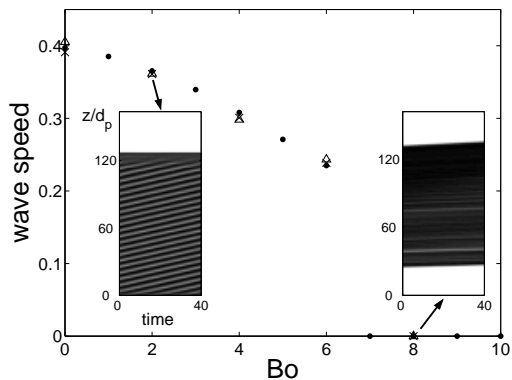}
\end{center}
\caption{\label{threecases}
Traveling wave speed (filled circles) in non-vibrated gas-fluidized
beds ($\Gamma = 0$), for various levels of cohesion. $U_s = 0.12$.
Beds of larger $Bo$'s ($> \sim$ 6) do not get fluidized;
rather the whole bed moves up as a plug.
Insets are spatiotemporal plots of moderately cohesive bed that
is fluidized ($Bo = 2$) and more cohesive bed ($Bo = 8$) that
is not fluidized and rises as a plug at this flow rate.
Triangles: The wave speeds obtained in beds of
$\sim 5d_p \times 5d_p \times 200d_p$.
Crosses: The wave speeds obtained in beds of
$\sim 10d_p \times 10d_p \times 100d_p$.
}
\end{figure}

We simulate both non-vibrated and vibrated gas-fluidized
beds of non-cohesive or cohesive particles ($0 \leq Bo \leq 50$).
We consider beds of narrow square-shaped cross sectional area
$5d_p \times 5d_p$ or $10d_p \times 10d_p$ with static depth
$H_0 \sim 100d_p$ and $H_0 \sim 200d_p$ (which consist of
3000 and 6000 particles respectively, in beds of
$5d_p \times 5d_p$ cross section).
Periodic boundary conditions are imposed in both lateral
directions, in order to avoid strong side wall effects
in beds of such a small aspect ratio.
We check that our results do not sensitively depend on a
particular choice of cross sectional area or the depth
of the bed (e.g., see Fig.~\ref{threecases}). We mostly
use a bed of $\sim 5d_p \times 5d_p \times 100d_p$ in the
following computations, unless otherwise stated.
We used $d_p$, $1.5d_p$, and $2d_p$ for the grid spacing
$\Delta z$.
The detailed profiles of the coarse-grained variables slightly
depend on the choice of $\Delta z$ (the bigger the grid size
is, the smoother the variables are, as one can readily expect),
but the main results remain the same, unless the grid size is
too large;
we set $\Delta z = 1.5d_p$ in all the results presented here.
In the following, all the quantities will be shown in
non-dimensional form.

\subsection{Bubbling and one-dimensional traveling waves}
 \label{1DTW}

We start by considering conventional (non-vibrated) gas fluidized
beds of non-cohesive particles to make certain that our model
captures basic experimental observations.
We first estimate the minimum fluidization velocity through
a simulation of quasi-static increase in the gas flow rate
(from zero) and the measurement of the pressure drop across
the bed, which yields $U_{mf} \approx 0.022$.
This estimate is slightly smaller than what we can compute
using the force balance relation and the approximate formula
of Wen and Yu, $U_{mf} \approx 0.023$~\cite{davidson85};
in this calculation, we used $\phi_{mf} = 0.652$, which
is measured from the bulk of the bed.
When the gas flow rate exceeds $U_{mf}$, the bed in our model
starts to expand inhomogeneously, and forms alternating bands
of plugs and voids.
Particles located at the bottom of one plug ``rain down''
through a void region and accumulate at the top of the lower
plug, causing the void regions to rise to the top in a
periodic fashion (Fig.~\ref{Usthree}).
This phenomenon corresponds to the formation of a periodic
train of bubble-like voids in real fluidized beds,
which appears as 1D-TW in the narrow beds we consider.
These waves represent the first stage in the bifurcation hierarchy
leading to various inhomogeneous structures in higher-dimensional
(i.e. 2D or 3D) fluidized beds.~\cite{glasser96,glasser98,sundar03}
The secondary bifurcations which occur in real fluidized beds
are suppressed in our simulations, which are forced to retain
the 1D character.
As the gas flow rate increases, both the wave speed and amplitude
increase (Fig.~\ref{Usthree}).
In the subsequent Sections, we will use the 1D-TW as an
indicator that characterizes the fluidizability of the bed.

\begin{figure}[t]
\begin{center}
\includegraphics[width=.8\columnwidth]{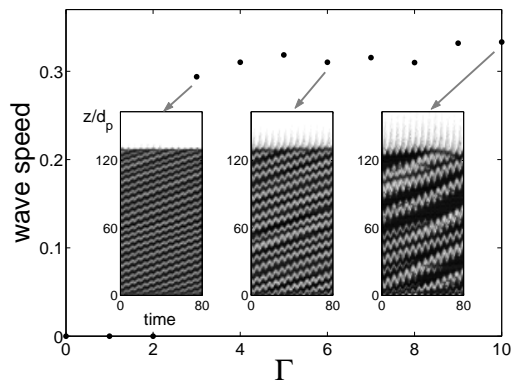}
\end{center}
\caption{\label{threecasesVib}
Traveling wave speed in beds of highly cohesive powders ($Bo = 20$)
subject to vibrations with various levels of $\Gamma$.
The traveling wave reappears for $\Gamma > 2$.
Insets are spatiotemporal plots of three different cases
($\Gamma = 3.0, 6.0, 10$), shown in the vibrating plate frame.
$f = 0.16$; $U_s = 0.12$.
}
\end{figure}

\subsection{Cohesion, vibration, and fluidization}
 \label{cohesion}

When the gas flow rate is well above $U_{mf}$, the bed exhibits
clearly visible 1D-TW [for instance, see the cases of $U_s = 0.082$
and 0.12 in Fig.~\ref{Usthree}].
In this Section, we examine the influence of the cohesion (between
particles) on the fluidizability of a bed, and explore how mechanical
vibration facilitates the fluidization of beds of cohesive particles.

Figure~\ref{threecases} shows the effect of $Bo$ on the wave speed,
where we have kept $St \sim 55$ and $U_s = 0.12$ (the same values
as in the last panel of Fig.~\ref{Usthree}).
As $Bo$ increases, the wave gradually slows down, and eventually
disappears.
At $Bo = 6$, the wave travels intermittently, remaining stationary
for some time and traveling at other times, and the wave speed
during the non-stationary phase is shown in Fig.~\ref{threecases}.
For $Bo > 7$, the whole bed rises up as a plug at this flow
rate, which is consistent with the well-known experimental
observations in narrow beds of strongly cohesive particles
(see a review article by Sundaresan~\cite{sundar03} and
references therein).
Further increase of $U_s$ only slightly improves the
fluidizability of the bed, confirming that beds of highly
cohesive particles cannot be fluidized by simply increasing
the gas velocity.

The results shown in Fig.~\ref{threecases} are obtained with
the assumption that the interaction between the particle
{\em and the base plate} is non-cohesive.
We check that the adhesion at the base plate does not make
any difference in the above results, as well as all the main
results presented here.
The only noticeable difference is that strongly cohesive
particles at the bottom of a bed get stuck to the plate for
some time (during a cycle; in vibrated beds), depending on
the oscillation parameters and the superficial gas flow velocity.
Detailed comparison between the two cases (with and without
adhesion) is shown later in Fig.~\ref{platestress}.

\begin{figure}[t]
\begin{center}
\includegraphics[width=.8\columnwidth]{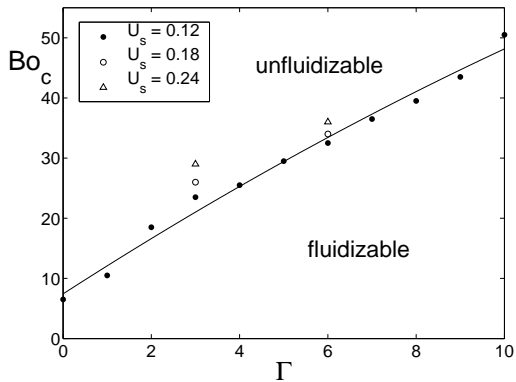}
\end{center}
\caption{\label{criticalBoG}
The critical values of $Bo$, above which the bed does not get
fluidized at given vibration parameters, as a function of $\Gamma$.
Three different levels of $U_s$ are used for $\Gamma =$ 3.0
and 6.0. $f = 0.16$.
A solid line, the least square fit for the cases of $U_s = 0.12$,
is included to guide the eye.
}
\end{figure}

\begin{figure}[b]
\begin{center}
\includegraphics[width=.8\columnwidth]{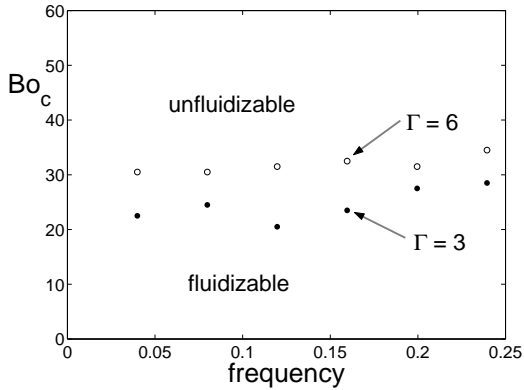}
\end{center}
\caption{\label{criticalBof}
The critical values of $Bo$ as a function of $f$,
for two different levels of $\Gamma$.
$U_s = 0.12$.
}
\end{figure}

Now we subject beds of even more cohesive particles ($Bo = 20$)
to mechanical vibration of a single frequency sinusoidal
oscillation in the direction of gravity.
When the vibration intensity is strong enough (when $\Gamma$
exceeds a certain value), even these highly cohesive beds get
fluidized in the sense that 1D-TW reappears (Fig.~\ref{threecasesVib}).
At a fixed gas flow rate (we assume it to be time-independent;
see the discussion in Sec.~\ref{1d-bed}),
the wavelength apparently increases with $\Gamma$; however,
the wave speed remains nearly the same (Fig.~\ref{threecasesVib}).
We define the critical Bond number $Bo_c$ as the maximum value
of $Bo$ for which the bed is fluidizable (generating 1D-TW) at
given set of oscillation parameters, and compute it as functions of
$\Gamma$ and $f$, using bisection-type search along the $Bo$-axis.
We find that $Bo_c$ increases almost linearly with $\Gamma$
(Fig.~\ref{criticalBoG}), but it only weakly depends on
$f$ (Fig.~\ref{criticalBof}).
The frequency range shown in this figure corresponds to the
usual operation range between 10 Hz and 100 Hz,
for the particle size of $d_p = 50~\mu$m.
In this range of frequencies, $Bo_c$ is virtually independent
of the frequency.
As the gas flow rate increases, $Bo_c$ slightly increases at
fixed values of $\Gamma$ and $f$ (see the cases for $\Gamma$
= 3 and 6 in Fig.~\ref{criticalBoG}).

\subsection{The role of vibration: Direct impact vs. pressure pulsation}
 \label{vibration}

In this Section, we discuss the manner in which the vibrating
base plate or the distributor, interacts with the bed material.

In the absence of gas, the kinetic energy of individual
particles in vibrated beds is obtained only from direct impact
with the plate, and dissipated through interaction between
particles.
In shallow beds ($H_0 <\sim 20d_p$), the fluctuating kinetic
energy (granular temperature) dissipates so quickly through
collisions that the whole bed can be well approximated by one
solid body.
The dynamics of the center of mass of such a bed can be
described by that of single perfectly inelastic ball on
a vibrating plate.~\cite{mehta90}
Vibrated shallow beds in vacuum undergo period doubling
bifurcations as $\Gamma$ is varied.~\cite{moon02}
Such layers of a
large aspect ratio form various spatiotemporal standing wave
patterns.~\cite{melo94,umbanhowar96}
However, temporal dynamics of vibrated deep beds (of about 20
particle deep or more) in vacuum are not commensurate with the
oscillation frequency, and exhibit more complicated non-periodic
behavior.~\cite{umbanhowar}

\begin{figure}[t]
\begin{center}
\includegraphics[width=.8\columnwidth]{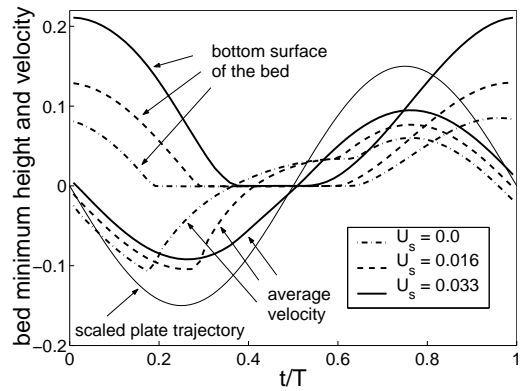}
\end{center}
\caption{\label{bed_dynamics}
The bottom surface and the average vertical velocity of vibrated
beds (averaged over all the particles in the bed) during a cycle,
shown in the {\em vibrating frame},
obtained for three different gas flow velocities.
$Bo = 20$; $\Gamma = 3.0$; $f = 0.16$.
A thin solid line, an arbitrarily rescaled plate trajectory
in the laboratory frame, is drawn to represent the phase
angle during a cycle.
}
\end{figure}

When the gas phase effects are accounted for, the gas drag
causes the dynamics of vibrated deep beds to deviate from
those in vacuum, and the deviation is more pronounced for
smaller particles (because the gas drag is larger).
For deep beds of fine powders considered here, the presence
of gas phase modifies the bed dynamics so that the temporal
dynamics become periodic.
In the absence of a net flow ($U_s = 0$) the bed lifts off
from the plate only slightly (even smaller than the particle size)
during a fraction of a cycle, and the bed impacts the plate later
in the same cycle (dot-dashed lines in Fig.~\ref{bed_dynamics}).
This periodic behavior has the same periodicity as the plate
oscillation.
Even for higher values of $\Gamma$, period doubling phenomenon,
which occurs in vibrated shallow layers in vacuum, is not observed.
Note that, in the early phase of the oscillation cycle shown in
Fig.~\ref{bed_dynamics} ($0 < t/T < 0.25$), the plate moves
downwards (and so does the base plate), and yet the bottom surface
of the bed is approaching the distributor plate.
i.e. the bed is descending faster than the plate.
The bed hits the plate, stays in contact for a duration of
time, and then detaches from it.

\begin{figure}[t]
\begin{center}
\includegraphics[width=.7\columnwidth]{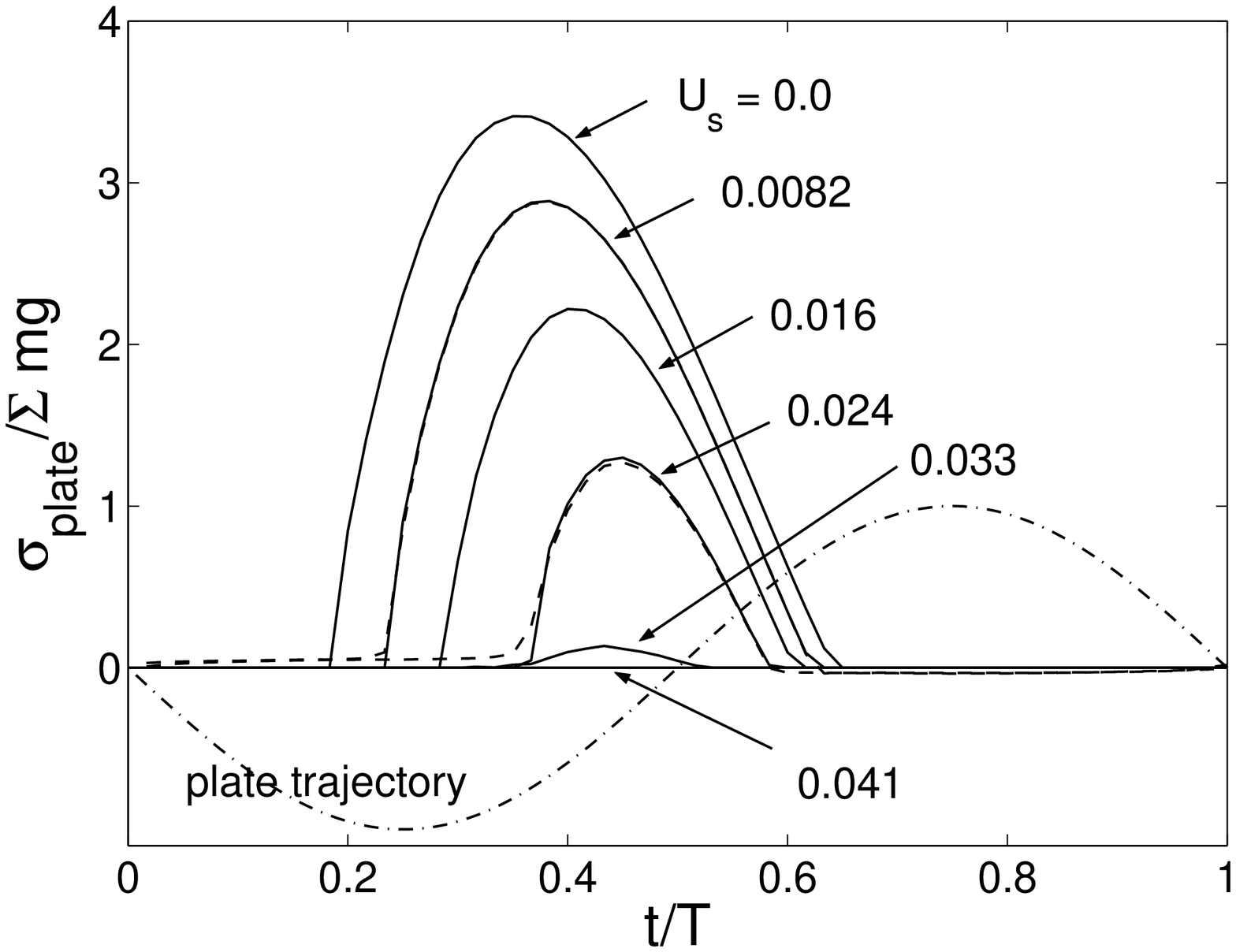}
\includegraphics[width=.7\columnwidth]{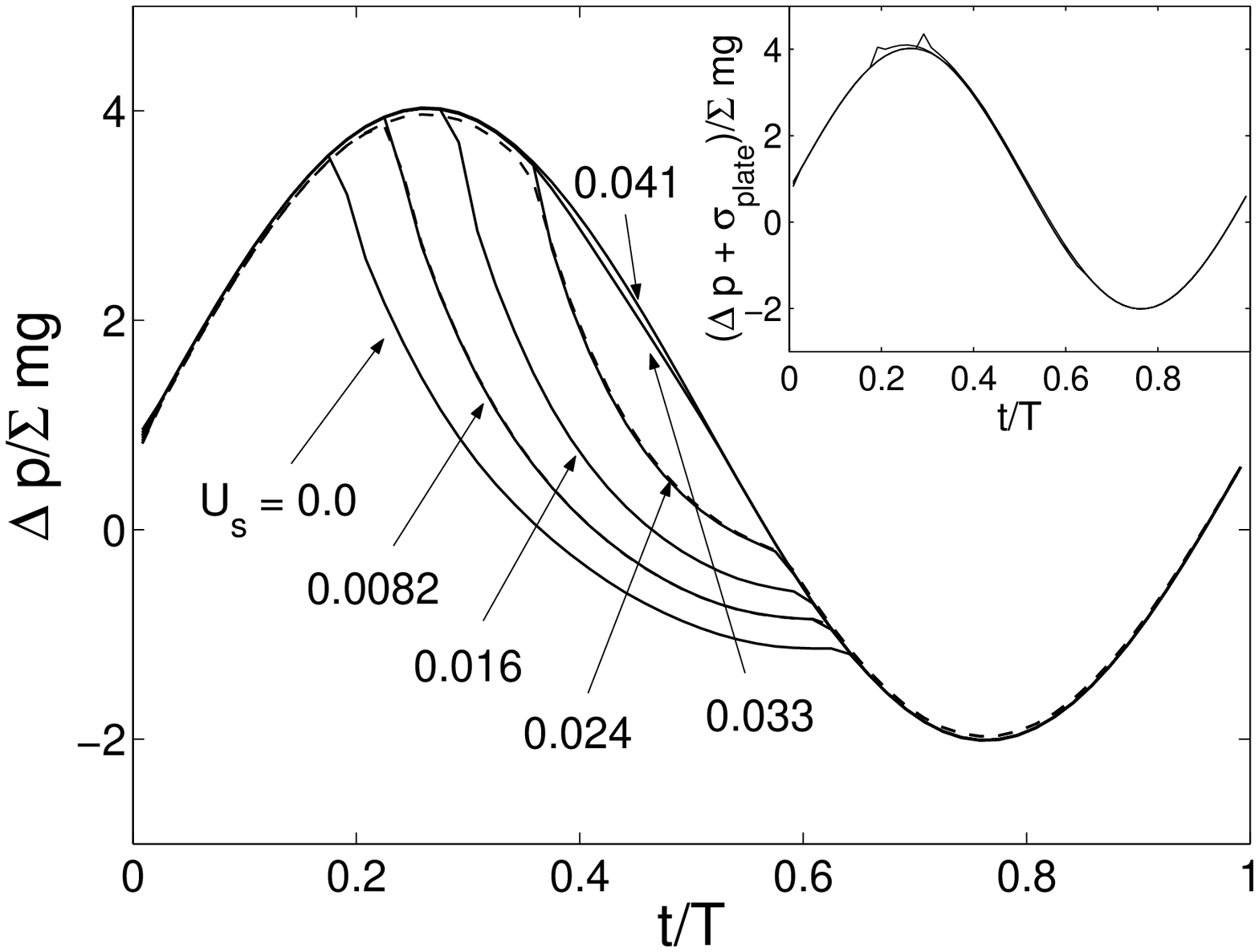}
\end{center}
\caption{\label{platestress}
(Top)
Normal stress tensor on the base plate,
scaled by the bed weight per unit cross sectional area
($\Sigma mg$), is shown during a cycle for different superficial
gas flow rates. $Bo = 20$; $\Gamma = 3.0$; $f = 0.16$;
$k_n = 2.0\times 10^5$.
Solid lines are obtained by neglecting adhesion at the base
plate ($Bo_p$, the $Bo$ at the plate, is 0).
Dashed lines are obtained when the adhesion is accounted for
($Bo_p = 40$), for two cases ($U_s = 0.0082, 0.024$), which
are nearly the same as the non-adhesive cases.
A dot-dashed line, the plate trajectory in an arbitrary unit,
is drawn to represent the phase angle during a cycle.
(Bottom)
The scaled pressure drop for the same cases as in the top panel.
Inset shows the sum of the pressure drop and the normal stress
at the base plate for three cases ($U_s = 0.0, 0.016, 0.033$),
all of which virtually coincide with the reduced effective
gravity ($g_{eff}/g$) in the vibrating frame
[sinusoidal curve; $1+\Gamma\sin(2\pi ft)$].
Small excess amounts, appearing as small peaks on top of
a sinusoidal curve, arise from the transient restoring force of
the {\em soft} particles at the impact, which gradually disappears
as the plate impacts the bed more gently with increasing $U_s$
(there are no apparent additional peaks for $U_s = 0.033$).
}
\end{figure}

When the gas flow is turned on and its rate gets increased,
the velocity of the bed (relative to the base plate) during
its short flight increases, and the bed lifts off further
from the plate (Fig.~\ref{bed_dynamics}).
The upward gas flow resists downward motion of the bed, hence
not only the duration of direct impact but also its magnitude
(strength) gradually decreases (top panel in Fig.~\ref{platestress}),
as the gas flow rate increases.
When the adhesion at the base plate is accounted for
(dashed lines in Fig.~\ref{platestress}),
the plate experiences some force even when the bulk of the
bed is in flight ($t/T < \sim 0.2$ or $> \sim 0.6$),
as there are a small number of particles stuck to the plate.
Other than this, compared to the case when the adhesion is
neglected (solid lines), no difference is observed.
The bed eventually hardly touches the plate at some gas
flow rate, above which direct impact remains minimal.
It will be shown in the next Section that the minimum gas
flow rate at which the direct impact virtually vanishes
is for all practical purposes the same as the minimum
fluidization velocity in vibrated beds.

\subsection{Pressure drop in vibrated beds}
 \label{pressuredrop}

The pressure drop across vibrated beds oscillates with
the same periodicity as the plate oscillation (bottom panel
in Fig.~\ref{platestress}).
The pressure drop increases (decreases) when the base plate
moves down (up); see Fig.~\ref{bed_dynamics} and the bottom
panel in Fig.~\ref{platestress}, for $t/T < \sim 0.5$.
At first glance, this seems to be counter-intuitive, as the
pressure drop increases when the plate is ``moving away''.
However, it should be noted that the change in the pressure
drop is determined by the change in the gap between the bed
and the base plate (i.e. the relative motion with respect to
the plate), not by the absolute motion of the base plate in
the laboratory frame.
As noted in the previous Section, the bed approaches the plate
during the phase of the oscillation cycle when the plate is
moving down from its mean position.

As soon as impact occurs, the pressure drop begins to decrease
rapidly, even below zero.
While the bed is moving away from the plate after the take-off
(see $t/T > \sim 0.6$ in Fig.~\ref{bed_dynamics}),
the pressure drop continues to decrease.
During this time, a region of lower pressure is being created
in the gap between the bed and the plate (bottom panel in
Fig.~\ref{platestress}).
Our results are generally consistent with the experimental
measurements by Thomas {\it et al.}~\cite{thomas00}; however,
direct comparison with their data is not possible because of
differences in systems and particle properties.
As the gas flow rate increases, the abrupt drop at the impact
gets smaller until it virtually vanishes
(and so does a sudden increase in the stress at the plate
$\sigma_{plate}$), and the pressure drop curve during a cycle
approaches nearly the same, asymptotic sinusoidal curve.

\begin{figure}[t]
\begin{center}
\includegraphics[width=.7\columnwidth]{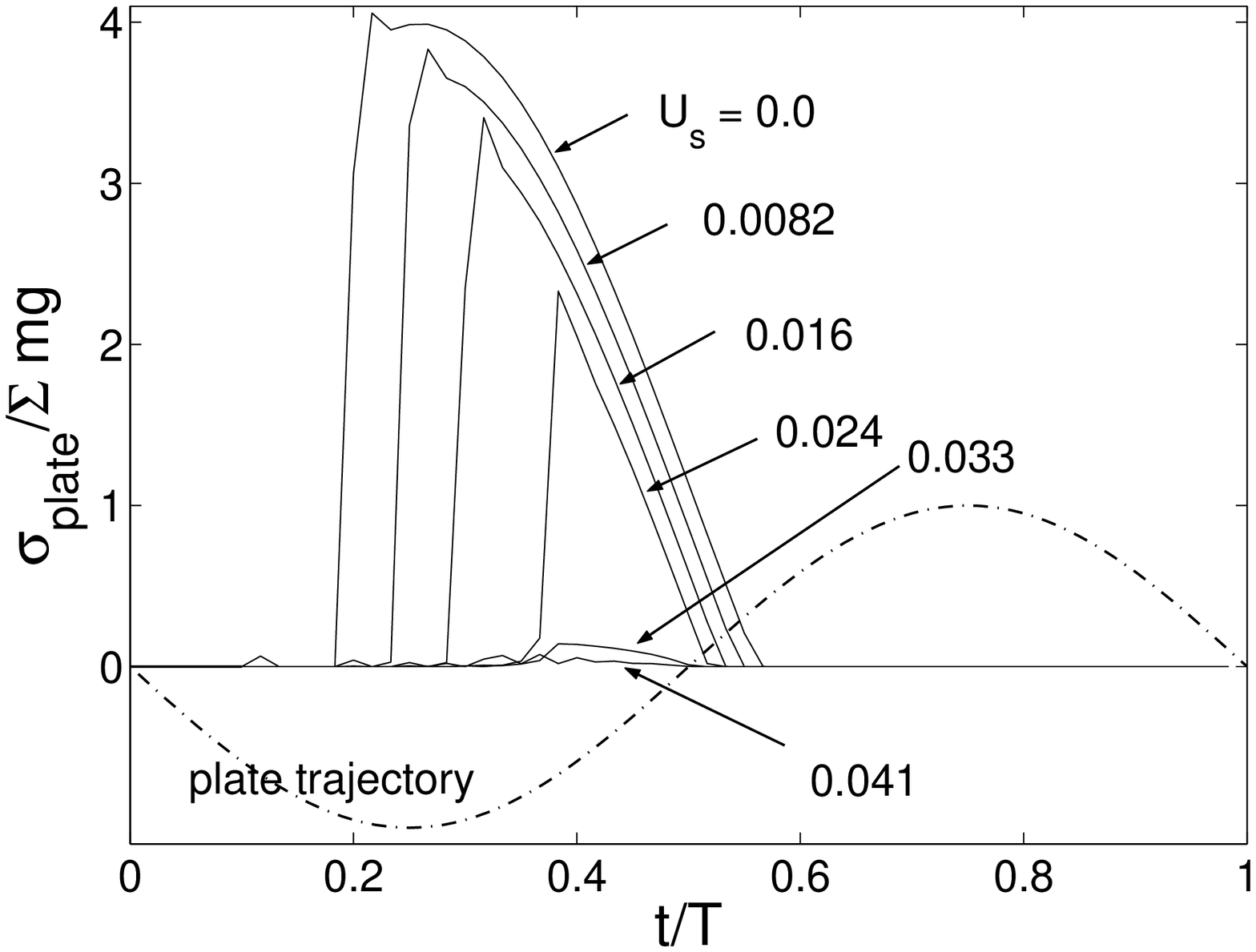}
\includegraphics[width=.7\columnwidth]{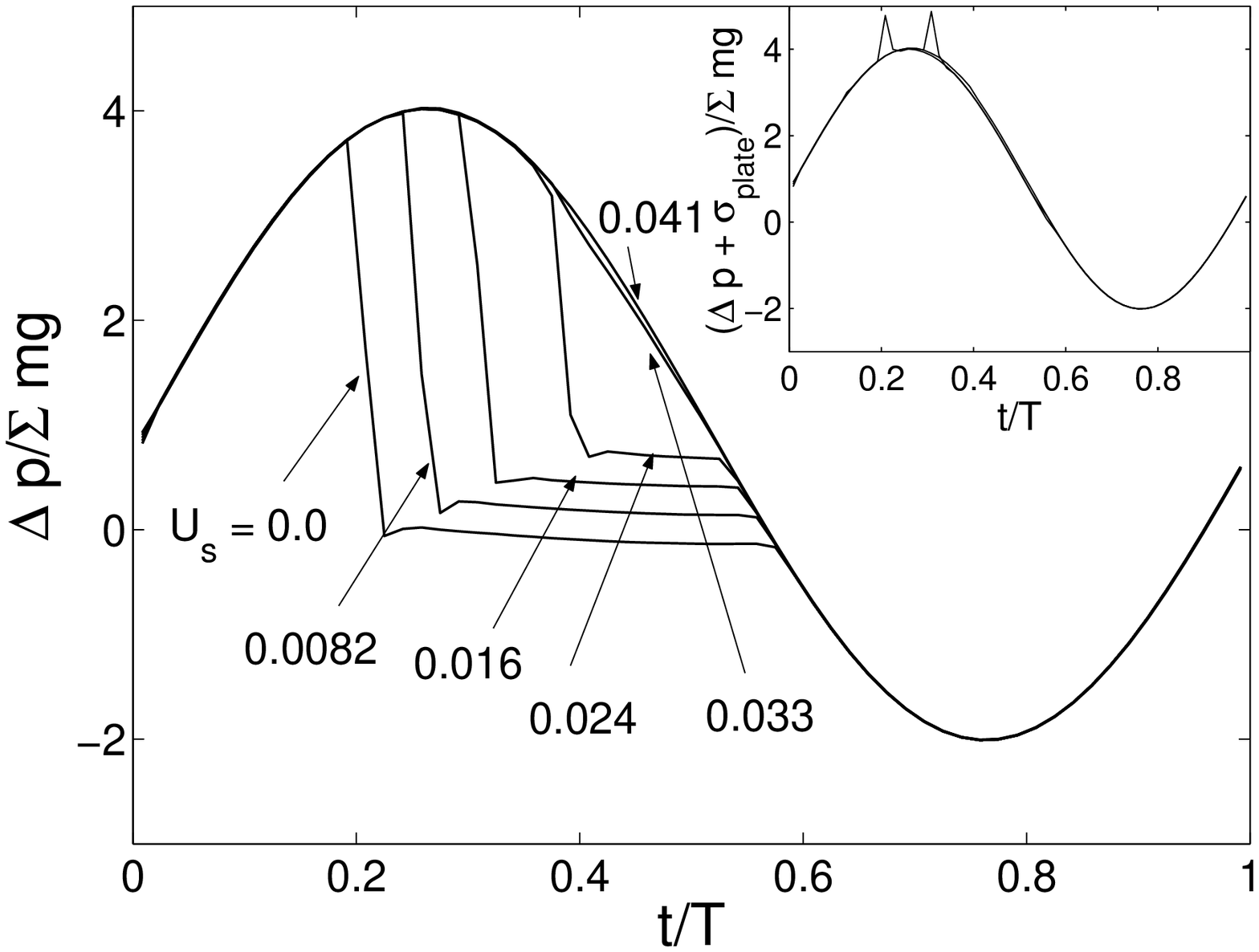}
\end{center}
\caption{\label{platestress2}
The same quantities measured for the same cases as in
Fig.~\ref{platestress} except that the spring stiffness
$k_n$ and $k_s$ have been increased by a factor of 10
(e.g., $k_n = 2.0\times 10^6$).
}
\end{figure}

When the vertical component of the stress at the plate
is combined with the pressure drop for each case, the resulting
curves for different gas flow rates virtually coincide with
the same sinusoidal curve
(inset in bottom panel of Fig.~\ref{platestress}).
This can be understood from the following force balance
relation in the direction of gravity:
\begin{equation}
 \label{forcebalance}
\frac{\sigma_{plate}}{\sum mg} + \frac{\Delta p}{\sum mg} = \frac{g_{eff}}{g},
\end{equation}
where $\sum mg$ is the weight of the bed per unit cross sectional
area, and the restoring force due to the softness of the particles
(i.e., the spring stiffness) is neglected.
The above relation holds at every moment during a cycle,
and the asymptotic common sinusoidal curve in the inset is
identified to be $1 + \Gamma\sin(2\pi ft)$. It corresponds to
the effective gravity in the vibrating plate frame $g_{eff}/g$,
the right hand side of Eq.~(\ref{forcebalance}).
Using the fact that the pressure drop across the bed is
limited by $(1+\Gamma)\sum mg$ during a cycle, we can
examine the validity of the incompressibility assumption
for the gas phase that we use:
By making use of the equation of state for an ideal gas, one can
show that the ratio of the change in gas phase density
$\Delta \rho_g$ to its reference value $\rho_{ref}$ satisfies
$\Delta \rho_g/\rho_{ref} \approx \Delta p/P_{atm} \leq
(1+\Gamma)\sum mg/P_{atm}$, where $P_{atm}$ is the atmospheric
pressure.
For the beds of fine powders considered in our study,
$\sum mg/P_{atm} \sim {\cal O}(10^{-5})$, hence
$\Delta \rho/\rho_{ref} \ll 1$ and the assumption of the
incompressibility is valid at {\em every} moment during a cycle.

\begin{figure}[t]
\begin{center}
\includegraphics[width=.8\columnwidth]{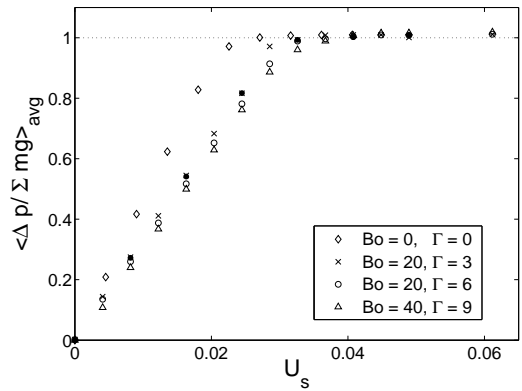}
\end{center}
\caption{\label{pdropAVG}
The average pressure drop during a cycle, scaled by the bed weight
per unit cross sectional area, obtained for different cases of
vibrated fluidized beds, as a function of $U_s$.
The pressure drop in conventional fluidized beds of non-cohesive
particles (diamonds) is included for comparison.
Filled circles are for the case of $Bo = 20, \Gamma = 3.0$, but
10 times larger value of spring stiffness ($k_n = 2.0\times 10^6$), 
%(both $k_n$ and $k_s$; i.e., the case in Fig.~\ref{platestress2}),
compared to all the other cases ($k_n = 2.0\times 10^5$), was used.
}
\end{figure}

In order to test the sensitivity of the results to the spring
stiffness, we repeated the calculations shown in
Fig.~\ref{platestress} with an order of magnitude larger value
of $k_n$. The results are presented in Fig.~\ref{platestress2}.
The detailed behavior during a cycle surely depends on the value
of the spring stiffness; the beds of softer particles in
Fig.~\ref{platestress} noticeably further compress and expand
during the impact, and the pressure drop keeps decreasing
even below zero {\em during} the impact.
This effect diminishes when a larger value is used for $k_n$;
in spite of the quantitative changes that are readily apparent,
the results shown in Figs.~\ref{platestress} and \ref{platestress2}
are qualitatively similar.
Importantly, in a fully fluidized state, pressure pulsation
is the only relevant mechanism in both cases, and the value
of $k_n$ becomes irrelevant.

Viewing the pressure pulsation at the bottom plate as a forcing
set up by the plate, one can inquire about the speed at which
this pulsation propagates upwards through the bed.
As we have taken the gas phase to be incompressible, the pulse
propagates nearly instantaneously. Thus, at each time instant,
the gas pressure decreases essentially monotonically as one moves up
through the bed (except for the small periodic variation associated
with the voidage waves).
If one allows the gas to be compressible, then the pulsation
travels at the speed of sound; sound speed through gas-fluidized
beds is considerably smaller than that through a column of
gas.~\cite{gregor75}
If the time required for the propagation of the pulsation through
the bed is commensurate with the period of the plate oscillation,
resonance can set in; however, such resonance is suppressed in
the present study as we have treated the gas as incompressible.
In any case, in the relatively shallow beds that we consider,
resonance is not expected to be a significant effect.

\begin{figure}[t]
\begin{center}
\includegraphics[width=.8\columnwidth]{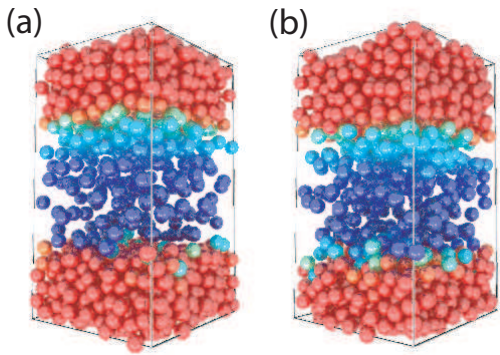}
\end{center}
\caption{\label{snapshots}
[Color online: Particles are color-coded according to the
vertical velocity, increasing from blue (moving downwards)
to green to red (moving upwards)]
Snapshots of beds of (a) non-cohesive particles ($Bo = 0$)
and (b) cohesive particles ($Bo = 4$), obtained from particles
in a box of height $L = 18d_p$ and cross sectional area
$10d_p \times 10d_p$ ($\phi_{avg} = 0.40$).
Periodic boundary conditions are imposed in all three
directions (see text).
Particles rain down from the upper plug to the lower plug,
while void regions in the middle rise up, which form traveling
voidage waves.
Note that cohesive particles form strings or agglomerates
while they rain down.
Full animations are available from
{\sf http://multiphase.princeton.edu/ICE\_2005}.
}
\end{figure}

The pressure drop averaged throughout a cycle as a function
of gas flow rate manifests a linear increase up to the constant
plateau region (Fig.~\ref{pdropAVG}), which is qualitatively
the same as in conventional fluidized beds.
The minimum velocity when the direct impact virtually does
not occur is essentially the same as the minimum fluidization
velocity in vibrated beds; only the pressure pulsation is
a relevant mechanism for a fluidized state of vibrated beds.
Note that the minimum fluidization velocities for the vibrated
beds of cohesive particles are larger than that for a
non-vibrated bed of non-cohesive particles of the same size.
This can be interpreted as the increment in effective
particle size, which is understandable, because the fluidized
entities in vibrated beds of cohesive particles are agglomerates,
not individual particles.
The average volume fraction at the minimum fluidization is smaller
(e.g. $\phi_{mf} = 0.631$ for the case of $Bo = 40$), compared to
what is observed in a bed of non-cohesive particles (0.652);
cohesive beds tend to pack more loosely.
In the fully fluidized state, the pressure drop exhibits a plateau,
which approximately equals the weight of the bed per unit
cross sectional area (Fig.~\ref{pdropAVG}), as in non-vibrated beds.
There is no clear consensus on this issue in experimental studies.
Tasirin and Anuar~\cite{tasirin01} found that the pressure drop
increases as the vibration intensity $\Gamma$ increases, Erd\'esz
and Mujumdar~\cite{erdesz86} observed the opposite trend, and
Marring et al.~\cite{marring94}, Mawatari et al.~\cite{mawatari02},
and Nam et al.~\cite{nam04} observed constant plateau pressure drop
equaled with the bed weight per unit cross sectional area
in high gas flow rates.
Only the latter is consistent with our results, which can be
explained by the simple force balance argument in
Eq.~(\ref{forcebalance}), accounting for the fact that the
direct impact is negligibly small in a fluidized state.

\subsection{Break-up of cohesive assembly by pressure pulsation}
 \label{breakup}

We seek to understand how vibration facilitates the break-up of
cohesive assemblies into agglomerates and maintains the propagation
of the wave in a fluidized state. We analyze the profile of
continuum variables, including the granular temperature $T$
and the solid phase stress tensor $\sigma$ (or the solid phase
pressure), across the traveling wave during a cycle.
As the waves in a bed of finite depth that we have considered
thus far are not perfectly periodic, we consider an alternate,
idealized geometry, where a wave is fully developed in a small
periodic box (in all of three directions) of height $L$ that is
commensurate with the wavelength obtained in the vibrated fluidized
bed simulations described above. Note that the direct impact does
not play an important role for a fully fluidized bed
(Sec.~\ref{vibration}; only the pressure pulsation does),
and the vibrating plate does not have to be considered in such
a case.
For a fully fluidized state, the weight of the bed per unit cross
sectional area is supported by the pressure drop:
\begin{equation}
 \label{3dperiodic}
p|_{z=0}-p|_{z=L} = \rho_p g_{eff} \phi_{avg} L,
\end{equation}
where $\phi_{avg}$ is the average volume fraction.

Comparison between fully fluidized states of cohesive beds and
non-cohesive beds on microscopic level reveals that cohesive particles
form strings of particles or agglomerates while they rain down
through void regions, whereas non-cohesive particles come down
individually (Fig.~\ref{snapshots}; full animations are available
from {\sf http://multiphase.princeton.edu/ICE\_2005}).
As one can readily see, this effect arises from the attractive
force between cohesive particles, which can be well characterized
by tensile stress on a continuum level.
We compute the stress tensor, consisting of a kinetic or dynamic
part and a virial or static part, using the following
microscopic relation~\cite{latzel00}:
\begin{equation}
\mathbf{\sigma} = \frac{1}{V} \left[\sum_i m_i \mathbf{\widetilde{v}}_i
\otimes \mathbf{\widetilde{v}}_i - \sum_{c \in V} \mathbf{f}_c
\otimes \mathbf{l}_c \right ],
\end{equation}
where $\mathbf{\widetilde{v}}_i = \mathbf{v}_i - <\mathbf{v}_i>$
is the fluctuating velocity of the $i$th particle,
$\otimes$ is the dyadic tensor product,
$\mathbf{f}_c$ is the interacting force between contacting particles
1 and 2, and $\mathbf{l}_c = \mathbf{r}_{1} - \mathbf{r}_{2}$
is the displacement vector between the centers of particles
under consideration.
The second term is summed over all the contacts in the averaging
volume $V$.

\begin{figure*}[t]
\begin{center}
\includegraphics[width=1.4\columnwidth]{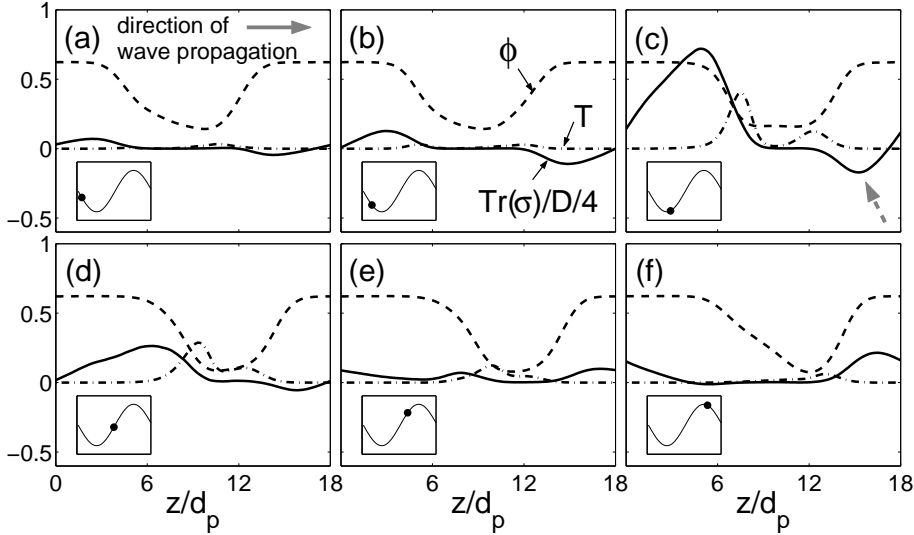}
\end{center}
\caption{\label{onecycle}
Solid phase continuum variables of a highly cohesive bed
($Bo = 20, \phi_{avg} = 0.40$) during a cycle in the vibrating
frame ($\Gamma = 3.0$), obtained in the same geometry as in
Fig.~\ref{snapshots}.
Insets represent the phase angle during a cycle in terms of the
vertical location of the (virtually existing) base plate
(indicated by dots).
The void region in the middle oscillates up and down, and has
an upward net motion [indicated by a gray arrow in panel (a)].
The solid phase stress during a cycle can be summarized by the
following stages:
(a) The pressure pulsation is weak and the particle phase pressure
is negligibly small throughout the bed.
(b) As the pressure pulsation increases, so do the magnitude of
the normal stress both in compressive (in lower plug) and tensile
regions (in upper plug) and granular temperature,
(c) until they reach their maximum values; around this time
the tensile stress reaches the tensile strength of the material,
which breaks up the cohesive assembly into agglomerates.
(d) Then the stress relaxes.
(e) and (f) Relatively small compressive stress builds up at the
bottom of the upper plug, but this is not relevant for the break-up
of the assembly.
The location of the peak tensile stress is indicated by a broken
arrow in panel (c).
A full animation during two cycles is available from
{\sf http://multiphase.princeton.edu/ICE\_2005}.
}
\end{figure*}

Now we consider the continuum level interpretation of vibrated
fluidized beds of highly cohesive particles in a fully periodic
box during an oscillation cycle (Fig.~\ref{onecycle}).
During a cycle, the wave oscillates up and down with a net
upward motion, as can be seen in cases of Fig.~\ref{threecasesVib}.
We compute the solid phase continuum variables including
the pressure, the trace of the stress tensor per dimension
Tr$(\sigma)/D$.
We compute them in the co-traveling frame (with the wave's net
motion), averaged over 100 cycles for the purpose of variance
reduction.
While the pressure pulsation cyclically varies during a cycle,
so do both the granular temperature and the stress tensor.

Figure~\ref{onecycle} shows the variation of solid phase continuum
variables at six different times during an oscillation cycle.
The inset in each panel shows the position of a hypothetical,
oscillating base plate at the instant the profiles of the
continuum variables are shown.
The plate trajectory sketched in the insets can be compared to
that shown in the top panel of Fig.~\ref{platestress}.
The corresponding pressure drop across the periodic domain can
be found from the cyclically varying pressure drop profile in
the bottom panel of Fig.~\ref{platestress}, which is nearly
out of phase with the plate position.
In Fig.~\ref{onecycle} (a), $g_{eff}/g$ and the scaled pressure
drop are nearly unity; they are larger in Fig.~\ref{onecycle} (b);
attain their largest values near the instant shown in
Fig.~\ref{onecycle} (c); relax back towards unity
in Fig.~\ref{onecycle} (d); and then they decrease in
Fig.~\ref{onecycle} (e) until they reach the minimum values
in Fig.~\ref{onecycle} (f).

When the scaled pressure drop is nearly unity [Fig.~\ref{onecycle} (a)],
the solid phase pressure is negligibly small throughout the bed.
As the pressure drop increases until it reaches its maximum
value [Figs.~\ref{onecycle} (b) and (c)], the granular temperature
and compressive stress in the lower plug and tensile stress
in the upper plug also increase significantly until they reach
their maximum magnitudes.
The large granular temperature occurs at the lower plug, where
the particles or agglomerates get accumulated.
It arises from vigorous collisions among raining down particles
or agglomerates and the lower plug.
Note that the volume fraction in this region is still low
($\phi <\sim 0.3$).
A rough estimate for the magnitude of the stress using the
values at the maximum temperature region
%($\sim nT = 6/\pi \phi T \sim 2\phi T \sim 0.2$), where the
($\sim nT = 6/(\pi d_p^3) \phi T \sim 0.2$), where the
most vigorous collisions occur, shows that it is still far
smaller than the tensile strength of the cohesive assembly
($\sim 1$) that is estimated below;
here $n$ is the number density.
In order for these vigorous collisions to contribute to the break-up
of the cohesive assembly, the magnitude of the stress formed
by the collisions has to be comparable or larger than the
tensile strength of the material.
However, the stresses formed by the collisions are not strong
enough to break up the assembly, and irrelevant for the assembly
break-up.
Rather, it is the increased tensile stress in the upper plug
that breaks up the assembly into agglomerates and maintains
the wave propagation.
Particles in the upper plug can split off from the assembly and
rain down, because the increased tensile stress becomes large
enough to reach the tensile strength of the cohesive assembly.
It occurs at a time around that shown in Fig.~\ref{onecycle} (c).

As the magnitude of the tensile stress cannot exceed the
strength of the assembly, we estimate the strength of
the fluidized bed by measuring the maximum tensile stress.
We compare such obtained tensile strength of the fluidized bed
with the prediction of Rumpf's model,~\cite{rumpf62}
\begin{equation}
\sigma_t = \frac{1-\epsilon}{\pi}k\frac{F_t}{d_p},
\end{equation}
where $\sigma_t$ is the tensile strength, $\epsilon$ is the porosity
($= 1 - \phi$), $k$ is the coordination number,
and $F_t$ is the cohesive contact force.
We find that our measurement ($\sigma_t \sim 0.8$) is about an order
of magnitude smaller than the prediction ($\sigma_t \sim 6$).
This discrepancy is understandable, because the cohesive assembly
in a fluidized bed breaks up through the weakest linkage,
as opposed to all directions as is assumed in the Rumpf's model.

Later in the cycle, when pressure drop decreases and approaches
the minimum value [Figs.~\ref{onecycle} (d) through (f)], the
particle phase pressure is generally small in the bed.
When the pressure drop is negative, relatively small compressive
stress builds up at the bottom of the upper plug as the gas
in the void region pushes up; this is obviously irrelevant for
the break-up.

\section{Summary}
 \label{conclusion}

We have used a particle dynamics-based model for vibrated
gas-fluidized beds of fine powders to study how the vibration
facilitates the fluidization of beds of cohesive powders.
We have demonstrated that, as the gas flow rate increases,
the direct impact from the plate decreases and the pressure
pulsation becomes more dominant.
In a fluidized state, the latter is shown to be virtually
the only relevant mechanism, and the pressure drop follows
a simple sinusoidal curve during a cycle (Figs.~\ref{platestress}
and \ref{platestress2}), which corresponds to the weight
of the bed per unit cross sectional area in the vibrated frame.
As a consequence, the pressure drop averaged over a cycle
in the fluidized state is simply the offset of this
sinusoidal curve, which equals the weight of the bed
per unit cross sectional area, as in non-vibrated beds.
This relation can be readily understood by a simple force
balance argument [Eq.~(\ref{forcebalance})].
In a bubbling bed (which appears as 1D-TW in our study),
it is during the transient time interval with large enough
pressure pulsation when the increased tensile stress breaks
up cohesive assembly into agglomerates (Fig.~\ref{onecycle}).

Note that the compressibility of the gas phase was ignored in
the present study; it would be interesting to investigate
resonance effects which may arise in vibrated beds by allowing
for gas compressibility.
It was also assumed that the gas superficial velocity is
independent of time throughout the oscillation cycle, and this
corresponds to the limit of very large resistance for gas flow
through the distributor;
it would also be interesting to examine the case of finite
distributor resistance, where the superficial gas velocity
can be expected to vary cyclically with plate vibration.

While the present study has yielded physical understanding of
the pressure fluctuations induced by the plate and the tensile
stress in the particle assembly, an analytical relation between
vibration parameters and the agglomerate size is still elusive.
Furthermore, we have considered only beds of narrow cross
sectional areas, and assumed the volume-averaged gas phase to
be 1D.
Consequently, some generic behavior in real vibrated gas
fluidized beds, such as horizontal sloshing motion of the
particles, meandering gas flows around agglomerates, and more
complicated bubble dynamics, were not allowed to occur.
By avoiding such complexity, we were able to bring out certain
basic physics of the bed dynamics.
Future studies should investigate the dynamics of higher
dimensional beds; however, simulation of fluidized beds of
realistic industrial scales, using the current approach,
is not yet feasible.

\section*{Acknowledgment}

This work was funded by The New Jersey Commission on Science
and Technology, The Merck \& Co., Inc., and an NSF/ITR grant.
We are delighted to contribute this article to the special
issue honoring WBR.
Having him as a colleague and a friend has been a pleasure
and a privilege for both SS and IGK.
We salute his accomplishments and look forward to many more
years of work and fun together.

\section*{Nomenclatures}
 \label{notation}

\begin{tabular}{cl}
$A$ & = Hamaker constant \\
$Bo$ & = cohesive Bond number between particles \\
$Bo_p$ & = cohesive Bond number between particles and base plate \\
$C_D$ & = drag coefficient \\
$\Delta p$ & = pressure drop across the bed \\
$\Delta z$ & = grid spacing for coarse-grained variables \\
$d_p$ & = particle diameter \\
$e$ & = normal coefficient of restitution \\
$f$ & = vibration frequency \\
$\mathbf{F}_{cont}$ & = interaction force due to contact \\
$\mathbf{F}_c$ & = cohesive force due to van der Waals force \\
$g_{eff}$ & = effective gravitational acceleration \\
$h$ & = halo function \\
$k$ & = coordination number \\
$k_n$ & = spring stiffness in the normal direction \\
$k_t$ & = spring stiffness in the tangential direction \\
$m_p$ & = mass of individual particle \\
$n$ & = solid phase number density \\
$P_{atm}$ & = atmospheric pressure \\
$p$ & = gas phase pressure \\
$Re_g$ & = Reynolds number based on particle size \\
$r$ & = particle radius \\
$s$ & = inter-surface distance \\
$St$ & = Stokes number \\
$T$ & = granular temperature \\
$\sum mg$ & = bed weight per unit cross sectional area \\
$\mathbf{u}_g$ & = volume-averaged gas phase velocity \\
$\mathbf{u}_s$ & = volume-averaged solid phase velocity \\
$U_s$ & = superficial gas flow velocity \\
$U_{mf}$ & = minimum fluidization velocity \\
$v_n$ & = relative velocity in normal direction \\
$v_s$ & = relative velocity in tangential direction \\
$\mathbf{v}_p$ & = velocity of individual particle \\
$V_p$ & = volume of individual particle \\
\end{tabular}

\section*{Greek symbols}
 \label{greek}

\begin{tabular}{cl}
$\beta$ & = inter-phase momentum transfer \\
$\delta,\delta^*$ & = minimum separation distance for cohesion \\
$\Delta_s$ & = tangential displacement from initial contact \\
$\Delta_n$ & = amount of overlap in the normal direction \\
$\epsilon$ & = porosity (gas phase volume fraction) \\
$\phi$ & = solid phase volume fraction \\
$\gamma_n$ & = damping coefficient for dashpot \\
$\Gamma$ & = vibration intensity \\
$\mu$ & = coefficient of friction \\
$\mu_g$ & = gas phase viscosity \\
$\nu_P$ & = Poisson's ratio \\
$\rho_g$ & = gas phase mass density \\
$\rho_s$ & = solid phase mass density \\
$\sigma$ & = solid phase stress tensor \\
$\sigma_t$ & = tensile strength of the material \\
$\sigma_{plate}$ & = vertical stress on the plate \\
\end{tabular}

\begin{table}[h]
\begin{center}
\caption{\label{table}
Parameter values used.
% ({\sf THIS IS INCOMPLETE; I WILL ADD MORE}).
}
%\begin{ruledtabular}
\begin{tabular}{|c|c|c|}
\hline
 & Typical values for & Nominal values for \\ 
 & dimensional quantities & dimensionless parameters \\ \hline
$\mu_g$ & 1.8$\times 10^{-4}$ g/(m$\cdot$s) & \\
$g$ & 981 cm/s$^2$ & \\
$d_p$ & 50 $\mu$m &\\
$\rho_s$ & 0.90 g/cm$^3$ & \\
$\sqrt{gd_p}$ & 2.2 cm/s & \\
$\sqrt{d_p/g}$ & 2.3$\times 10^{-3}$ s &  \\
$\delta$ & 0.4 nm ($= \delta^*$) & 8.0$\times 10^{-6}$ \\
$\Delta t$ & 5.6$\times 10^{-7}$ s & 2.5$\times 10^{-4}$ \\
$\Delta z/d_p$ & & 1.5 \\
$e$ & & 0.9 \\
$\mu$ & & 0.1 \\
$k_n$ & & 2.0$\times 10^5$ - 2.0$\times 10^6$ \\
$\nu_P$ & & 0.3 \\
$\Gamma$ & & 0 - 10 \\
$f$ & $< \sim$ 100 Hz& 0 - 0.25 \\
\hline
\end{tabular}
%\end{ruledtabular}
\end{center}
\end{table}

\end{document}